**Hydrogen Effect on Plastic Deformation and Fracture in Austenitic Stainless Steel**

Mr Eugene Ogosi
University of Aberdeen
King's College
Aberdeen, AB24 3FX
United Kingdom

Dr Amir Siddiq
University of Aberdeen
King's College
Aberdeen, AB24 3FX
United Kingdom

Dr Umair Bin Asim
University of Aberdeen
King's College
Aberdeen, AB24 3FX
United Kingdom

Dr Mehmet Kartal
University of Aberdeen
King's College
Aberdeen, AB24 3FX
United Kingdom

## ABSTRACT

The effect of hydrogen on the fracture behaviour of austenitic stainless steel has been investigated in the past [1][2]. It has been reported that fracture initiates by void formation at inclusions and regions of enhanced strain localisation [3]. There is experimental evidence that supports the fact that hydrogen influences void nucleation, growth and coalescence during material fracture [4]. This work investigates the effect of hydrogen on void growth and coalescence in austenitic stainless steel. The effect of hydrogen on void growth and coalescence for different stress triaxialities has been examined by analysing the stress strain response of a single crystal representative volume element (RVE). The results show that the higher the stress triaxiality, the lower the equivalent stress required to yield. This response is found to be similar irrespective of whether the material is being exposed to hydrogen or not. Lower equivalent strain values to yield were experienced for higher stress triaxialities for both hydrogen free and hydrogenated samples. Hydrogen slowed down void growth at high stress triaxialities but promoted void growth as lower triaxialities. For lower triaxialities, the presence of hydrogen was found to initially inhibit void growth at low equivalent strain values. However, this effect reversed at higher equivalent strain values and hydrogen was found to promote void growth. The effect of hydrogen promoting or inhibiting void growth have been shown to increase in magnitude with increasing hydrogen concentration.

Key words: Hydrogen Embrittlement, Stress Corrosion Cracking, Deformation, Plasticity, void growth

## 1. INTRODUCTION

Austenitic stainless steels have widespread application in the energy, nuclear, automobile, chemical, oil and gas production, refining and medical industries showing a superior strength range, ductility and corrosion resistance when compared to other types of steels [5]. In comparison to most steels, austenitic stainless steels have been used for a variety of hydrogen process and transport applications due to its high resistance to hydrogen related degradation [6]. However, austenitic stainless steels are vulnerable to Stress Corrosion Cracking (SCC) in specific environmental conditions including when exposed to hydrogen [1]. When steel is embrittled and fails by cracking due to exposure to hydrogen in the presence of stress, the failure mechanism is known as Hydrogen Induced Stress Corrosion Cracking (HISCC) [7]. Plastic deformation has been found to precede cracking in many failure cases for austenitic stainless steel and there has been a lot of commentary on the Hydrogen Enhanced Localised Plasticity (HELP) mechanism which has been used to explain this phenomenon [8][9][10]. Hydrogen Induced Stress Corrosion Cracking (HISCC) manifestation using the traditional fracture stages of void nucleation, growth and coalescence have also been invoked by several authors. There are many experimental and theoretical evidence to support this phenomenon. S.P Lynch [1] has provided a concise review of the various theories  and the reader is referred to this document for more details. In literature, the fracture of metals due to hydrogen can been classed into four main mechanisms; Hydrogen Enhanced Decohesion (HEDE), Hydrogen Enhanced Strain Induced Vacancy (HESIV), Hydrogen Enhanced Localised Plasticity (HELP) and embrittlement due to hydride formation [3][11].  The HELP and HESIV mechanisms are relevant here due to the observed changes in deformation and void behavior in the presence of hydrogen. HELP is a popular mechanism which is supported by experimental evidence and is used to explain the effect of hydrogen on plastic deformation of steel prior to and during fracture [2]. The basic premise of this theory is that atomic hydrogen screens elastic energy between dislocations and causes strain localisation leading to embrittlement and eventual fracture [2]. Another manifestation of HELP is the restriction of the flow of dislocation due to the presence of hydrogen [3]. HESIV is used to explain how hydrogen promotes fracture by affecting the traditional stages of fracture i.e. void nucleation, growth and coalescence [3]. HESIV proposes that hydrogen promotes strain localisation and this leads to vacancy formation, agglomeration to form voids, void growth and coalescence [3]. Experimental evidence have shown that hydrogen enhanced void nucleation may occur at sites of impurity (e.g. carbides) [12], however, impurities are not essential for void formation and voids have been observed to form at sites where no impurities exist [13]. These voids have been found to initiate in areas of high dislocation density and on deformation induced dislocation boundaries [14]. Yagodzinskyy et al [15] has provided experimental evidence that show hydrogen promotes dislocation – dislocation interaction which contributes to void formation during plastic deformation. Martin et al have previously demonstrated that the formation and extension of voids occurred along slip bands and were facilitated by the presence of hydrogen [16][17][18]. Bullen et al [19] introduced hydrogen into interstitial sites and observed that hydrogen increased void density in face centred cubic nickel material but void size was reduced. Jiang et al investigated the effects of hydrogen on void nucleation on 310 stainless steel. They found that hydrogen induced void formation by stabilising micro voids and decreased void surface energy [20]. Chen et al found experimentally that hydrogen-induced cracking happened through nano-void nucleation and then quasi-cleavage [21]. The density, size and morphology of voids are very useful in the prediction of the role hydrogen plays in this process of fracture [22]. Atomic hydrogen have been observed to slow down void growth with experiments performed for austenitic stainless steels and other FCC metals exposed to hydrogen [23][24]. Other researchers have observed that hydrogen promotes void nucleation (increased void density), growth and coalescence [3][19][25]. The failure stage which proceeds via the void initiation, growth and coalescence process may transpire due to either or a combination of internal necking of the inter-void ligament, internal shearing of the inter-void ligament and link up void length [26]. Internal necking is driven by sufficient void growth which allows neighbouring voids to link up by deformed necking of void ligament. Internal shear failure results from lack of void growth leading to void link up via shearing of void ligament [27]. It has been observed by Matsuo et al [19] that unlike carbon steel (which has a body centred cubic structure), hydrogen induced internal shear localisation in austenitic stainless steel results in increased void density. Material specimens with dislocation trapping characteristics related to austenitic stainless steel were shown to fail predominantly by internal shear failure in the presence of hydrogen especially at low stress triaxiality. Hydrogen induced internal shearing failure occurs with no substantial void growth, so that fracture features on the specimens manifest as small dimples supporting the observation that void growth is impeded by atomic hydrogen in the trapping sites. The presence of hydrogen promotes plastic instabilities and reduces the stress at which shear localisation occurs [27].  Material fracture is believed to mostly occur at locations of strain localisation [28] and high dislocation density[28][29]. Localisation occurs either by shear banding and necking bifurcation. The term internal necking failure is used to refer to fracture brought about by necking bifurcation and the term internal shearing failure is used to refer to fracture brought about by shear banding [30].

The results from numerical modelling also support the phenomenon that void growth is enhanced by the presence of hydrogen [24]. A finite element analysis method was used by Liang et al to analyse the effect of void growth on a two-dimensional unit cell containing a circular void for Niobium. The model used is on a rate independent constitutive law that assumes that the material yields according to the von-Mises criterion, and hardens isotropically under plastic straining.. Hydrogen was found to have a softening effect on the material and void volume fraction increased in the presence of hydrogen. Ahn et al [31] simulated void growth and coalescence for steel using a finite element analysis software. A 2D unit cell containing a void was constructed and hydrogen effect on Laing's elastoplastic constitutive relationships [24]. Hydrogen was found to promote void growth and coalescence. At triaxiality =3, void coalescence was observed to take place by accelerated hydrogen induced strain localization around the void [31].

In our previous work, a crystal plasticity model was presented which related critical resolved shear stress to plastic slip due to dislocation motion and crystal strength [32]. The effects of hydrogen on plastic deformation and void growth was presented and validated by comparing with experimental results. Hydrogen was shown to increase the critical resolved shear stress and work hardening during stages I and II of plastic deformation. Preliminary results showed that void growth was slower in the presence of hydrogen. This paper presents further results on the effect of different hydrogen concentrations on void growth for different stress triaxialities. The theoretical basis of the model is presented in sections 2 and 3. The methodology used for this study is explained in section 4. Results are presented and discussed on void growth for various triaxialities in section 5. Conclusions and recommendations are presented in section 6

## 2.    CRYSTAL PLASTICITY THEORY

The theory is based on crystal plasticity and has been extended to account for the effects of hydrogen on plastic deformation. Elastoplastic deformation is assumed to be driven wholly by crystalline slip and dislocation motion. The kinematics for the constitutive theory has been obtained by multiplicative decomposition of the deformation gradient using a finite strain theory [33]

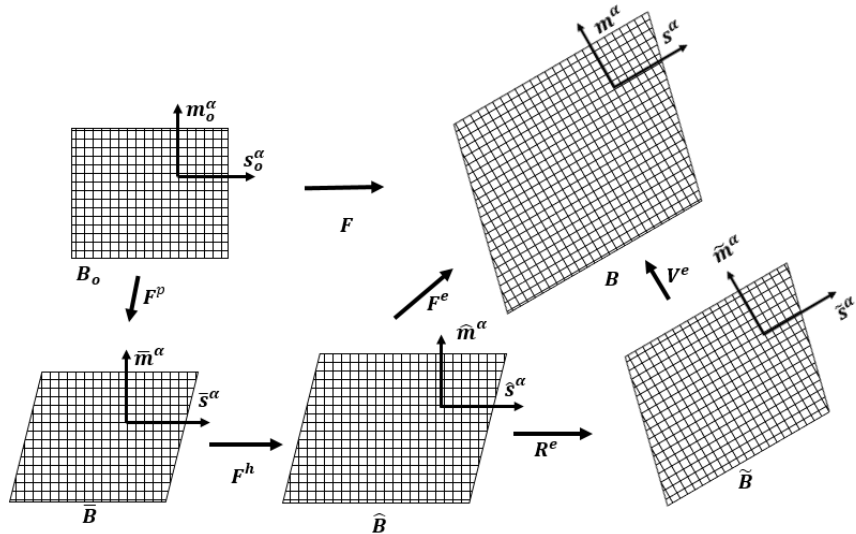

**Figure 1:  Multiplicative decomposition of the deformation gradient**

The deformed material illustrated in Figure 1 is assumed to be acted on by the vectors $\mathbf{x}^1$, $\mathbf{x}^2$, $\mathbf{x}^3$ and $\mathbf{x}^4$ which represent a set of particle positions in the configurations $\boldsymbol{B_o}$, $\overline{\boldsymbol{B}}$, $\widehat{\boldsymbol{B}}$ and $\boldsymbol{B}$ respectively. $\boldsymbol{B_o}$ is the initial configuration and $B$ represents the final state at time t after deformation. Hypothetically, unloading from the final state gives a configuration $\widetilde{B}$ that has the permanent plastic deformation and the dilatational distortion of the material induced by the presence of hydrogen. A second intermediate configuration $\bar{B}$ represents the state of the material due to plastic deformation without the hydrogen effect. $F^e, F^h$ and $F^p$ represent the elastic, hydrogen and plastic parts of the deformation gradient respectively.

$$F = \frac{\partial x_i^4}{\partial x_j^1}; \quad F^p = \frac{\partial x_i^2}{\partial x_j^1}; \quad F^h = \frac{\partial x_i^3}{\partial x_j^2}; \quad F^e = \frac{\partial x_i^4}{\partial x_j^3} \tag{1}$$

Applying the chain rule for partial differential equations gives;

$$\frac{\partial x_i^4}{\partial x_j^1} = \frac{\partial x_i^2}{\partial x_j^1} \frac{\partial x_i^3}{\partial x_j^2} \frac{\partial x_i^4}{\partial x_j^3} \tag{2}$$

or

$$F = F^e F^h F^p \tag{3}$$

Marin [34] crystal plasticity formulation is modified to include the effects of hydrogen;

$$F = V_e F^*, \quad F^* = R^e F^h F^p \tag{4}$$

$F$ represents the deformation gradient in the configuration $B$ from an original configuration $B_o$. $V_e$ is the elastic stretching. $R^e$ is the rigid body rotation, $F^p$ represents deformation gradient for the intermediate configuration $\bar{B}$. $F^h$ represents the deformation gradient for the configuration $\hat{B}$. The configuration $\widetilde{B}$ is obtained theoretically by unloading the elastic stretch from $B$ through $V^{e-1}$ without excluding $R^e$ and is represented by the deformation gradient $F^*$. The velocity gradient $l$ in $B$ is given as;

$$l = \dot{F}F^{-1} \tag{5}$$

$\dot{F}$ is the rate of change of the total deformation gradient and $F^{-1}$ is the inverse of the deformation gradient. For the $\widetilde{B}$ configuration the velocity gradient $(\tilde{L})$ is defined as

$$\tilde{L} = V^{e-1}lV^e = V^{e-1}V^e + \tilde{L}^* \tag{6}$$

$V_e$ is the elastic stretch tensor and $V^{e-1}$ is the inverse of the elastic stretch. Combining (4), (5) and (6) gives:

$$\tilde{L}^* = \dot{R}^e R^{eT} + R^e \hat{L}^h R^{eT} + R^e F^h \bar{L}^p F^{h-1} R^{eT} \tag{7}$$

$\tilde{L}^*$ is plastic flow due to slip, dilatational effect of hydrogen and rotation in $\widetilde{B}$. $R^e$ is the rotation tensor and $\dot{R}^e$ is the rate of change of the rotational vector. $\bar{L}^p$ is the velocity gradient due to plastic flow. $R^{eT}$ is the rotation tensor. $\hat{L}^h$ is the velocity gradient due to dilatational effect of hydrogen in the $\hat{B}$ configuration and is expressed by the Sofronis relationship [29]:

$$\hat{L}^h = \dot{F}^h.F^{h-1} = \frac{d}{dt}\left(1 + \frac{(c-c_o)\lambda}{3}\right)I.\left[\left(1 + \frac{(c-c_o)\lambda}{3}\right)\right]^{-1}I = \frac{1}{3}\Lambda(c)\dot{c}I, \quad \Lambda(c) = \frac{3\lambda}{3+(c-c_o)} \tag{8}$$

$\dot{F}^h$ is the rate of change of the total deformation gradient due to hydrogen and $F^{h-1}$ is the inverse of this tensor. $c$ and $c_o$ represent the current and initial concentrations of hydrogen respectively (expressed in hydrogen atoms per lattice atom). $\lambda$ is $\frac{\Delta V}{V_m}$, $\Delta V$ is the change in volume and $V_m$ is the mean atomic volume. $\dot{c}$ Is the rate of change of hydrogen concentration.

Diffusion of hydrogen in austenitic stainless steels is relatively slow [15] so the concentration of hydrogen during deformation at a certain material point during plastic deformation is assumed to be constant. Experimental evidence and justification of this assumption has previously been discussed by Schebler [35]. Based on this, the term "$c - c_o$" tends to zero and $L^h$ is reduced to an identity matrix. However, plastic deformation is affected by hydrogen

$\bar{L}^p$, is defined as:

$$\bar{L}^p = \sum_{\alpha=1}^{N} \dot{\gamma}^\alpha \ \bar{s}^\alpha \otimes \bar{m}^\alpha \tag{9}$$

$\dot{\gamma}^\alpha$ is shear strain rate due to slip, $\bar{s}^\alpha$ and $\bar{m}^\alpha$ are direction and normal to slip respectively. Substituting these results in  (7), gives:

$$\tilde{L}^* = \tilde{\Omega}^e + \hat{L}^h + \sum_{\alpha=1}^{N} \dot{\gamma}^\alpha \ \tilde{s}^\alpha \otimes \tilde{m}^\alpha \tag{10}$$

$\tilde{\Omega}^e = \dot{R}^e R^{eT}$ representing rigid body elastic spin. $\tilde{s}^\alpha$ is $R^e \bar{s}^\alpha$ and $\tilde{m}^\alpha$ is $R^e \bar{m}^\alpha$. And $\hat{L}^h$ being a scalar matrix remains unaffected by the transformation.

The Second Piola-Kirchhoff stress, $\tilde{S}$, is given as:

$$\tilde{S} = \tilde{\mathbb{C}^e} : \tilde{E}^e \tag{11}$$

$\tilde{\mathbb{C}}^e$ is the anisotropic elasticity tensor and $\tilde{E}^e$ is the Green-Lagrange strain tensor.

Velocity gradient may be additionally decomposed into a symmetric, $d$, and skew, $w$, parts i.e. $l = d + w$.  In $\tilde{B}$, the rate of deformation tensor becomes:

$$\tilde{D} = V^{eT} d V^e = \dot{\tilde{E}}^e + \tilde{D}^* \tag{12}$$

$$\tilde{D}^* = \text{sym}(\tilde{C}^e \tilde{\Omega}^e) + \sum_{\alpha=1}^{N} \dot{\gamma}^\alpha \text{sym}(\tilde{C}^e \tilde{Z}^\alpha) \tag{13}$$

Spin is given as:

$$\widehat{W} = V^{eT} w V^e = \text{skew}(V^{eT} \dot{V}^e) + \widehat{W}^* \tag{14}$$

$$\widehat{W}^* = \text{skew}(\tilde{C}^e \tilde{\Omega}^e) + \sum_{\alpha=1}^{N} \dot{\gamma}^\alpha \text{skew}(\tilde{C}^e \tilde{Z}^\alpha) \tag{15}$$

where $\tilde{C}^e$ right Cauchy-Green tensor given as $\tilde{C}^e = R^e \tilde{C}^e R^{eT}$ and $\tilde{Z}^\alpha$ is the Schmid tensor expressed in $B$ given by $\tilde{Z}^\alpha = \tilde{s}^\alpha \otimes \tilde{m}^\alpha$.

Plastic slip evolution is defined as:

$$\dot{\gamma}^\alpha = \dot{\gamma}_0^\alpha \left[ \frac{|\tau^\alpha|}{\kappa_s^\alpha} \right]^{\frac{1}{m}} \text{sign}(\tau^\alpha) \tag{16}$$

$\dot{\gamma}^\alpha$ is shear strain rate on the $\alpha^{th}$ slip system, $\dot{\gamma}_0^\alpha$ is the reference shear strain rate on the $\alpha^{th}$ slip system, $\kappa_s^\alpha$ is the current slip system strength, $\tau^\alpha$ is resolved shear stress, $m$ is strain rate sensitivity parameter. Voce type hardening was incorporated in the model using the evolution relation in (17). The slip system is made to harden with the evolution of accumulated slip till a saturation value is reached, beyond which it deforms in a perfectly plastic manner.

$$\dot{\kappa}_S^\alpha = h_0 \left( \frac{\kappa_{S,S}^\alpha - \kappa_S^\alpha}{\kappa_{S,S}^\alpha - \kappa_{S,0}^\alpha} \right) \sum_{\alpha=1}^N |\dot{\gamma}^\alpha|, \quad \kappa_{S,S}^\alpha = \kappa_{S,S0}^\alpha \left[ \frac{\sum_\alpha |\dot{\gamma}^\alpha|}{\dot{\gamma}_{S0}} \right]^{1 \backslash m'} \tag{17}$$

$\dot{\kappa}_S^\alpha$ is the current rate of hardening, $\kappa_S^\alpha$ is the current strength of slip system, $h_0$ is the reference hardening coefficient, $\kappa_{S,S}^\alpha$ is the saturation value of strength which depends on the accumulated slip $\sum_\alpha |\dot{\gamma}^\alpha|$.

$\kappa_{S,0}^\alpha \quad \kappa_{S,S0}^\alpha, \dot{\gamma}_{S,0}^\alpha$ and $m'$ are the material parameters that define the plastic behavior of single crystal. Critical resolved shear stress (CRSS) of each of the slip systems is given as $\kappa_S^\alpha(t=0)$ in (17).

## 3. HYDROGEN EFFECTS

For over a century, hydrogen has been known to readily permeate the microstructure of metals [36] and negative effects of hydrogen on material properties have been observed [37][38]. Various authors have provided concise reviews and discussions of these effects [1][2][6]. In practical terms, hydrogen maybe introduced into steel during manufacturing, fabrication, in service or generated from chemical dissociation of water due to corrosion reactions [39]. Specifically for austenitic stainless steel, the effects of hydrogen on deformation and fracture properties have also been documented [40][41]. According to the HELP mechanism, hydrogen have been observed to have two competing effects on dislocation. Hydrogen either enhances the mobility of dislocation by screening elastic interactions [2][8] or inhibits dislocation motion [42]. The later phenomenon have been discussed in terms of hydrogen atoms "pinning" groups of moving dislocation [3]. It is assumed that hydrogen will either reside at normal interstitial lattice sites (NILS) or in trap sites and in accordance to the Oriani theory both of these sites remain in equilibrium [43] . The total hydrogen concentration $C_{Total}$ is given by the relationship:

$$C_{Total} = C_L + C_T$$

$C_L$ and $C_T$ are the concentration of hydrogen atoms residing in NILS and trap sites respectively. According to Fick's first law, there would be transfer between sites if a concentration gradient exists between these sites. The proportional relationship between the hydrogen activity in traps $a_T$ and NILS $a_L$ is captured by the equilibrium constant $K_T$

$$a_T = K_T a_L \tag{18}$$

Each site hydrogen activity $a_i$ relates to fraction of species occupancy $\theta_i$ as follows;

$$a_i = \frac{\theta_i}{1 - \theta_i} \tag{19}$$

Starting from a reference of $a_i = \theta_i$ and tending towards zero. Combining (18) and (19) gives

$$\frac{\theta_T}{1 - \theta_T} = K_T \frac{\theta_L}{1 - \theta_L} \tag{20}$$

$\theta_T$ is hydrogen occupancy of trap sites, $\theta_L$ is hydrogen occupancy of lattice sites and $K_T$ is equilibrium constant. Concentration of hydrogen residing in trapping sites $C_T$ is given as:

$$C_T = \theta_T \psi N_T \tag{21}$$

$\psi$ is the number of sites per trap and $N_T$ is the number of traps per unit lattice given as:

$$N_T = \frac{\sqrt{3}}{a_{fcc}} \rho \tag{22}$$

$a_{fcc}$ is lattice parameter for FCC metal. Evolution of bulk dislocation density $\rho$ is given as:

$$\int_0^t \dot{\rho} \, dt = (k_1 \sqrt{y}) \int_0^t \dot{\gamma}/dt \tag{23}$$

$\dot{\rho}$ represents incremental changes in dislocation density. $\dot{\gamma}$ represents incremental changes in strain. $k_1$ is a constant associated with immobilised dislocation and $\sqrt{y}$ is average spacing between dislocations [44]. This relationship considers the effects of hydrogen on material behaviour during stages I and II of deformation where there is a lesser influence of temperature and strain rate [45]. The total hydrogen concentration is assumed to be constant due to low hydrogen diffusivity in austenitic stainless steels [46]. We note that although total hydrogen concentration $C_{Total}$ is constant, transfer of hydrogen atoms from NILS to more energetically favourable traps created during plastic deformation occurs. Krom's relationship gives the relationship between $C_T$ and $N_T$ [47]

$$C_T = \frac{1}{2}\left[\frac{N_L}{K_T} + C_{Total} + N_T - \sqrt{\left(\frac{N_L}{K_T} + C_{Total} + N_T\right)^2 - 4N_T C_{Total}}\right] \tag{24}$$

$N_L$ is the number of atoms per unit NILS. We have previously introduced two terms to account for the changes to the material properties due to the presence of hydrogen; Hydrogen initial strength coefficient $H_i$ and the Hydrogen hardening coefficient $H_f$ that quantifies effect of hydrogen on strain hardening. Initial crystal strength $\kappa_{h,0}^{\alpha}$ in the presence of hydrogen is given by the following expression;

$$\kappa_{h,0}^{\alpha} = \kappa_{s,0}^{\alpha} * (1 + H_i C_{initial}) \tag{25}$$

$\kappa_{s,0}^{\alpha}$ is crystal strength in hydrogen free condition. $C_{initial}$ is the amount of hydrogen in trapping sites before plastic deformation and relates to hydrogen uptake given by Caskey Jr [48]

$$C_{initial} = f\,C_L e^{18400}/(RT) \tag{26}$$

$f$ is a fraction of alloy atoms associated with a unit length of dislocation. $C_L$ is concentration of hydrogen in lattice sites. 18400 J/mol is bonding energy for hydrogen to dislocations in austenitic stainless steels.

The evolution of crystal strength given in (17) is modified as follows:

$$\dot{\kappa}_s^{\alpha} = h_0 \left(\frac{\kappa_{s,S}^{\alpha} - \kappa_s^{\alpha}}{\kappa_{s,S}^{\alpha} - \kappa_{s,0}^{\alpha}}\right) \sum_{\alpha=1}^{N} |\dot{\gamma}^{\alpha}| (1 + H_f C_T) \tag{27}$$

## 4. METHODOLOGY

Finite Element (FE) Model simulations are done using ABAQUS/Standard analysis [49]. Three-dimensional representative volume element (RVE) models have been constructed and meshed using the ABAQUS/CAE with reduced-integrated, first-order linear brick elements (C3D8R). Simulations were based on displacement control test. Material response relationships and equations based on the crystal plasticity theory and hydrogen influence discussed in sections 2 and 3 have been implemented numerically through the user material (UMAT) subroutine in ABAQUS. FE simulations were performed by replicating experiments performed on an austenitic stainless steel single crystal subjected to uniaxial tension oriented for multi-slip. Experimental tensile tests referenced, were performed on AISI316LN austenitic stainless steel single crystals of nominal chemical composition 18Cr-12Ni-2Mo alloyed with 0.5 wt. % of nitrogen by Yagodzinskyy et al [3]. Samples were cut parallel to the (110) crystal plane and tensile loaded in the $\langle 001 \rangle$ direction at a strain rate of 8 x $10^{-4}s^{-1}$ with atomic hydrogen content of 0.64%. Figure 1 shows good agreement between FE results and experimental data. Model parameters identified using inverse modelling [50]–[52] are given in Table 1.

### Table 1 Identified Model Parameters

| Model Parameters | $H_i$ | $C_{initial}$(%at) | $K_1\sqrt{y}$ | $N_L/K_T$ | $c_{Total}$ (%at) | $H_f$ |
|---|---|---|---|---|---|---|
| | 1.6 | 0.05 | 2.4 | 5.5e+26 | 0.64 | 0.05 |

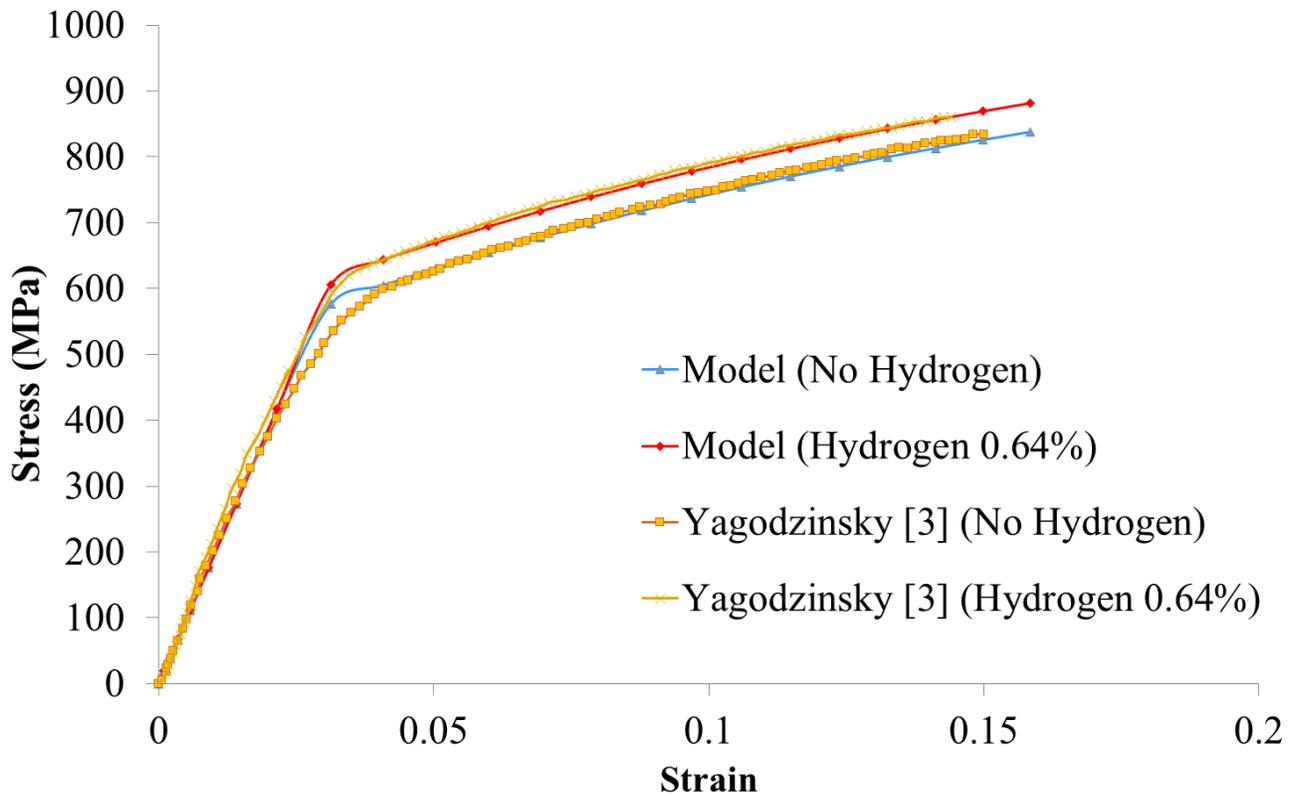

**Figure 1: Validated model results for AISI316LN austenitic stainless steel single crystals under uniaxial tensile loading**

For the analysis of fracture behaviour of the crystals at different stress states, an RVE model with an embedded void of known initial void fraction is constructed and meshed using ABAQUS/CAE with reduced-integration elements (C3D8R) (see illustration of void model in Figure 2).

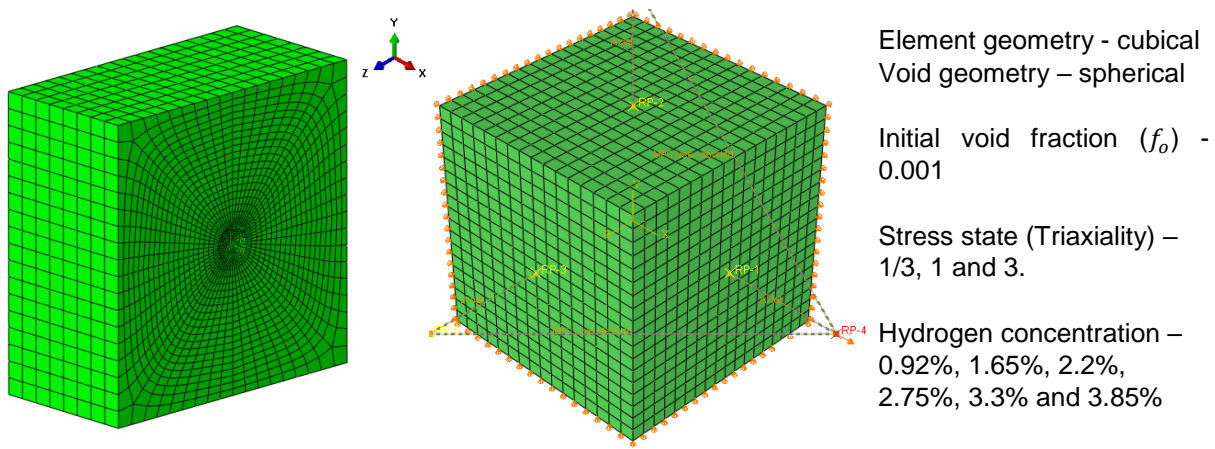

Element geometry - cubical
Void geometry – spherical

Initial void fraction ($f_o$) - 0.001

Stress state (Triaxiality) – 1/3, 1 and 3.

Hydrogen concentration – 0.92%, 1.65%, 2.2%, 2.75%, 3.3% and 3.85%

**Figure 2: RVE model with boundary conditions**

The porous crystal plasticity model and the relationship between void growth, strain, stress triaxiality, initial void size and crystal orientation have been discussed by other authors [53][54]. Void fraction evolution is defined as follows;

$$\text{Normalised void volume fraction} = \frac{f}{f_0}, \quad f = \frac{V_{void}}{V_{total}} \tag{28}$$

Where $f$ is void volume fraction, $f_0$ is initial void volume fraction, $V_{void}$ is the void volume and $V_{Total}$ is the total volume of the element i.e. solid material and void.

Using material parameters validated from experiments Table 1, void growth analyses were performed over selected ranges of hydrogen to observe the effects on the material stress strain response and analyse the effect of hydrogen on void growth. Displacements in the lateral direction were tuned to keep applied stress triaxialities constant through a multipoint constraint (MPC) user subroutine of the ABAQUS software, while volume averaged stress triaxiality was varied depending on the void growth.

## 5. RESULTS AND DISCUSSION

Figure 3 show that for the RVE sample with an embedded void of initial void fraction of 0.001, hydrogen was observed to increase the equivalent stresses and hardening response for triaxialities i.e. 1/3, 1 and 3. For triaxiality = 3, samples experienced an equivalent stress to yield value that was lower than that experienced by samples with triaxiality of 1/3 and 1. It was also observed that that the strain to fracture for triaxiality = 1/3 and 1 samples were higher than for triaxiality = 3. For all three stress states considered, the presence of hydrogen increased the stress and strain to yield. A similar response has been observed in finite element 2D modelling performed by Liang et al [24].

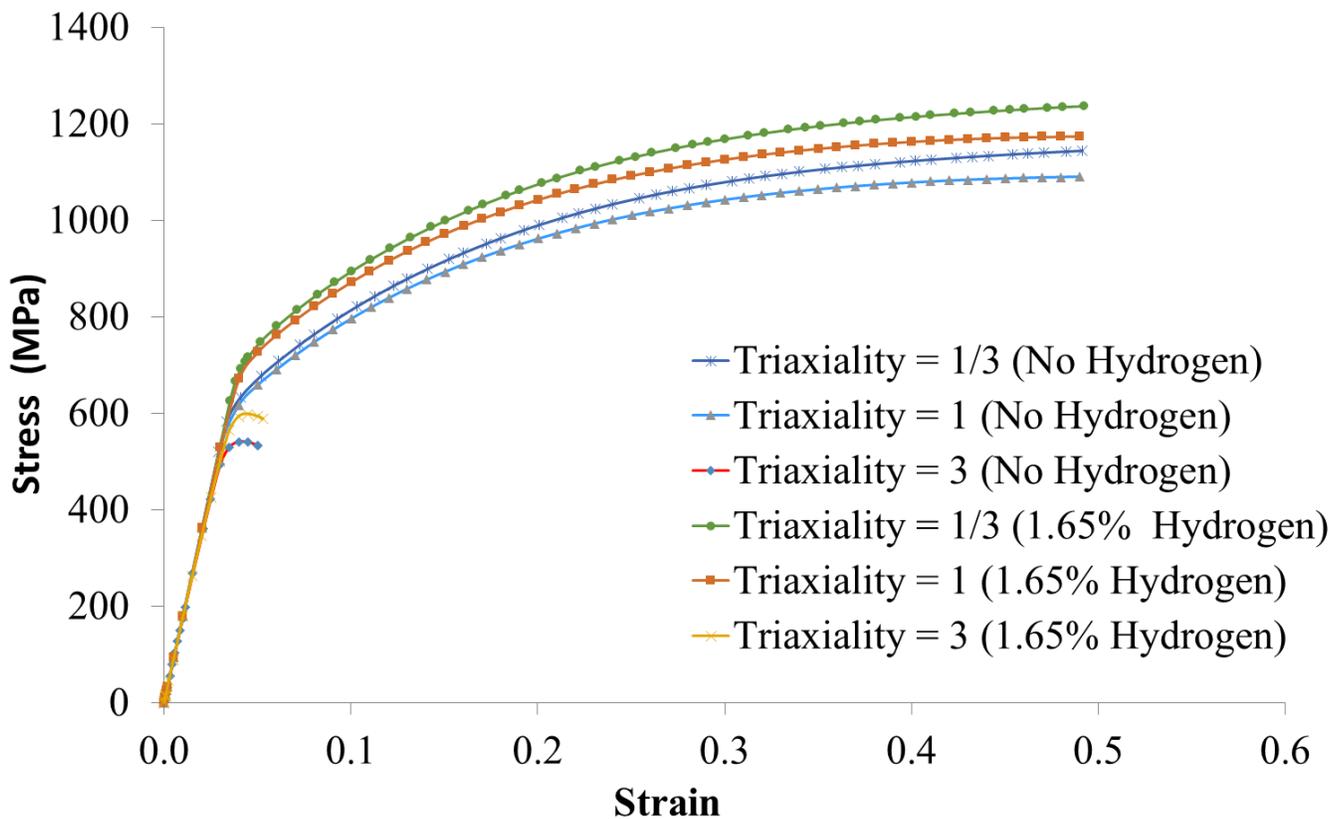

**Figure 3:** Hydrogen Effect on stress strain response at three different stress states, viz. uniaxial and triaxiality = 1 and 3, $f_0$=0.001

Figure 4 shows that void growth is higher at triaxiality = 3 when compared with triaxiality = 1. This was observed for both hydrogen containing and hydrogen free conditions. For all triaxialities, the void fraction is observed to first increase slowly and then rapidly which is consistent with previous findings [24][53][54]. The transition to an exponential increase in void fraction has been interpreted previously to be the onset of void coalescence [24]. It can also be observed from Figure 4 that for triaxiality = 3, the presence of hydrogen reduces void growth and delays the onset of void coalescence. For triaxiality = 1, the presence of hydrogen was observed to increase void growth and reduce the strain to void coalescence. However, the effect of hydrogen at smaller strain values for triaxiality 1 was found to be different. Hydrogen was observed to initially reduce void growth at strain values between 0.03 and 0.05 (refer marked area in **Figure 5** and zoomed view in Figure 6). However, as strain values increased further, the effect of hydrogen reversed and promoted void growth and reduced the strain to coalescence (Figure 6). This response was not observed for samples with triaxiality = 3 where strain values at coalescence were observed to be an order of magnitude smaller than for samples at triaxiality =1.

**Figure 8** and

**Figure 8** presents contour plots showing accumulated shear strain distribution at different equivalent strain values for triaxiality = 1 and 3. For both triaxialities, slip activity is found accumulate around the void and this is evidence that dislocations and microstructural defects are created around the voids. It is also observed that as equivalent strain increases, slip activity in the vicinity of the void increases. Figure 9 presents contour plots showing accumulated shear strain distribution for hydrogen concentration = 0% and 3.85% at an equivalent strain of 0.4 at stress triaxiality = 1. Figure 10 presents contour plots showing accumulated shear strain distribution for hydrogen concentration = 0% and 3.85% at an equivalent strain of 0.048 at stress triaxiality = 3. It can be seen from both figures that the presence of hydrogen increases the slip activity in the vicinity of the void for the same equivalent strain and triaxiality. The effect is also more noticeable for triaxiality = 3. This effect of hydrogen reducing the strain required for void coalescence at higher triaxialities have previously been noted and explained to be as result of a marked increase in dislocation activity in the area immediately bounding the void [31].

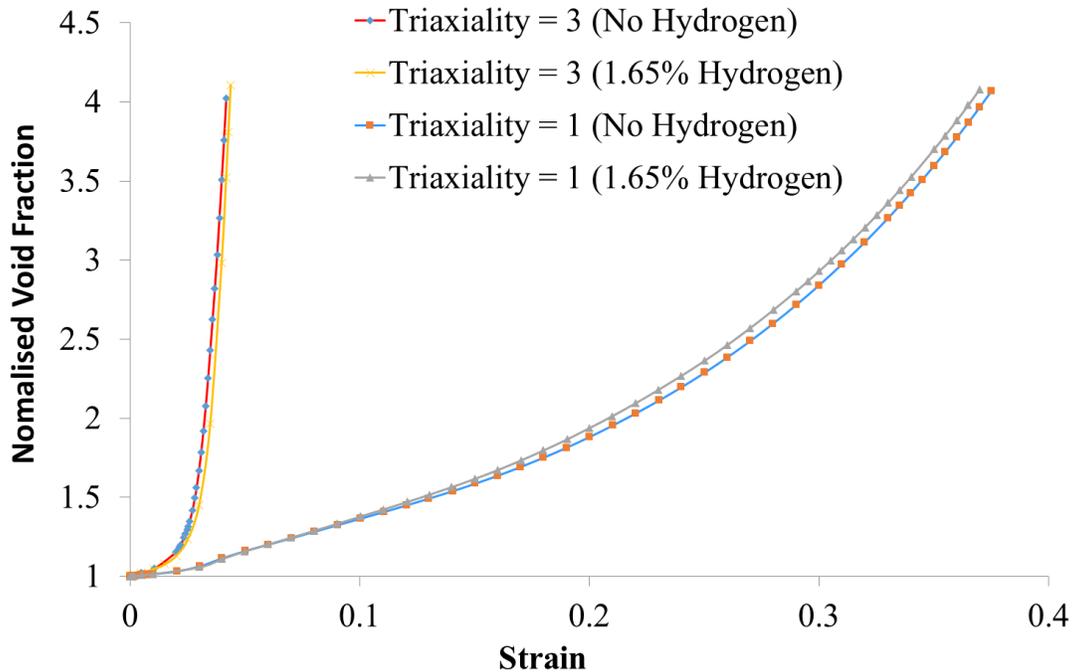

**Figure 4: Hydrogen Effect on Nominal Void Fraction at different Triaxialities**

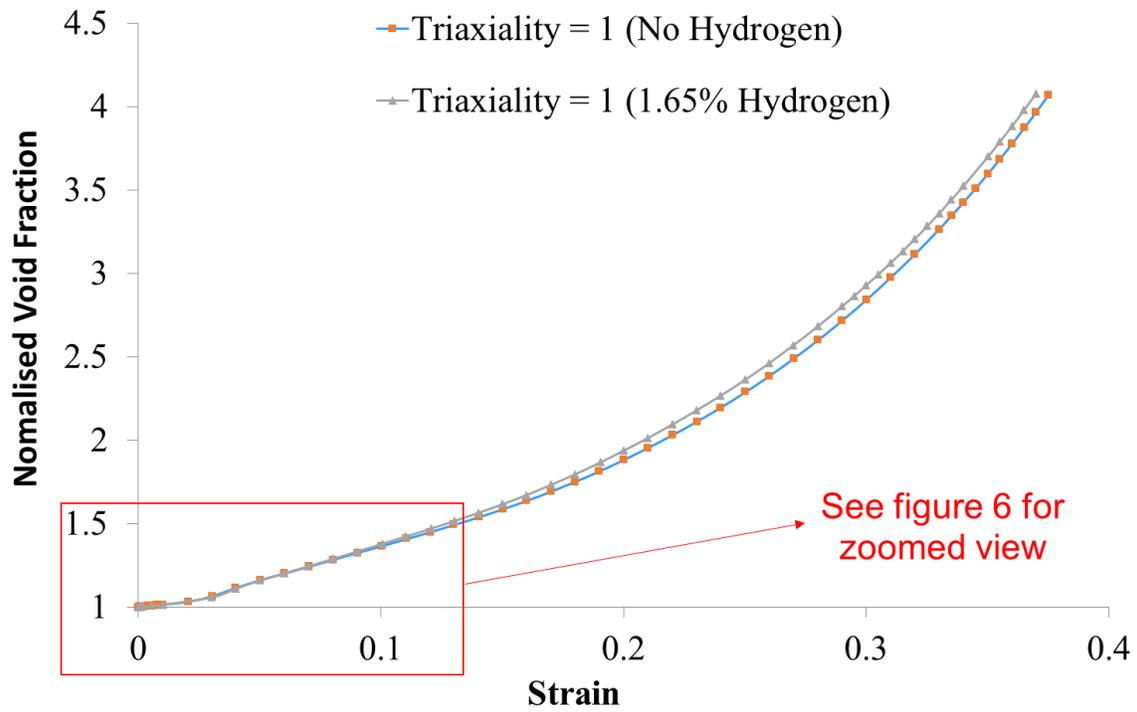

**Figure 5: Hydrogen Effect on Nominal Void Fraction at Triaxiality 1**

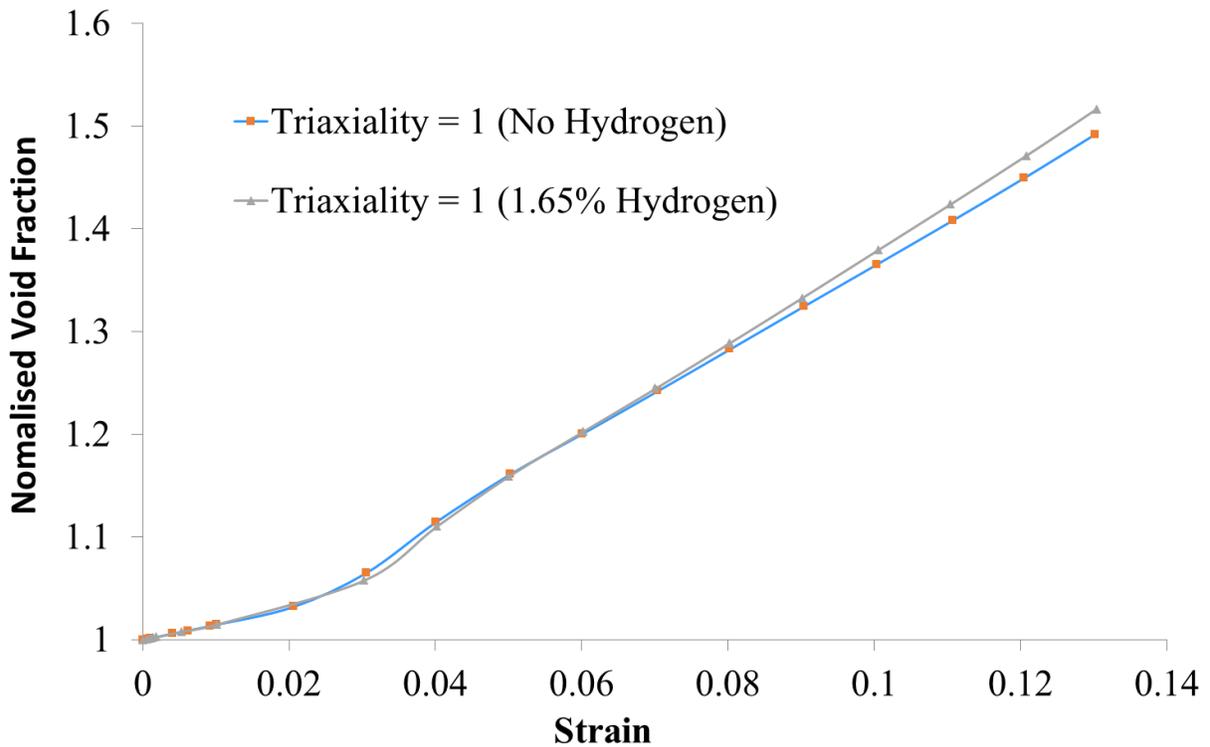

**Figure 6: Hydrogen Effect on Nominal Void Fraction at Triaxiality 1 (zoomed view)**

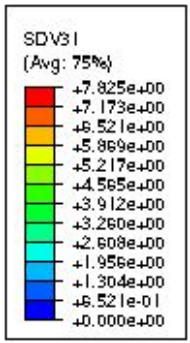

SDV 31 is accumulated
Plastic Sllip

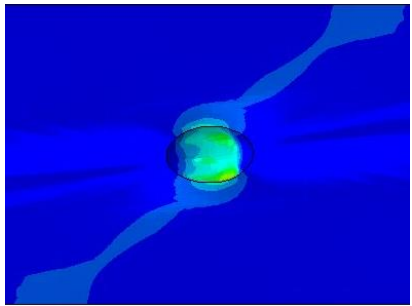

(a)  Equivalent Strain = 0.3

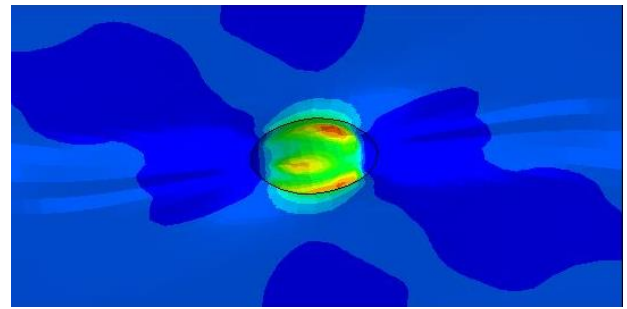

(b) Equivalent Strain = 0.43

**Figure 7: Contour Plots showing cumulative slip activity at different equivalent strains (a) Equivalent Strain = 0.3 and (b) Equivalent Strain = 0.4  Hydrogen Concentration = 3.85%, Triaxial State = 1 and initial void fraction = 0.001**

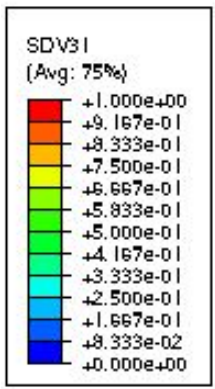

SDV 31 is Accumulated
Plastic Sllip

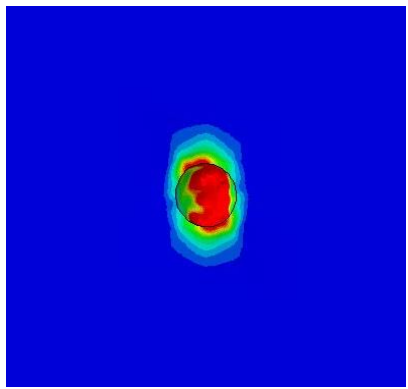

(a)  Equivalent Strain = 0.04

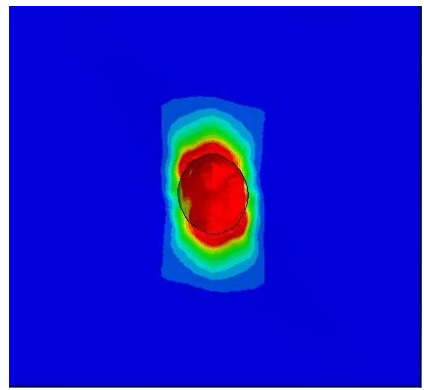

(b) Equivalent Strain = 0.048

**Figure 8: Contour Plots showing cumulative slip activity at different equivalent strains (a) Equivalent Strain = 0.04 and (b) Equivalent Strain = 0.05  Hydrogen Concentration = 3.85%, Triaxial State = 3 and initial void fraction = 0.001**

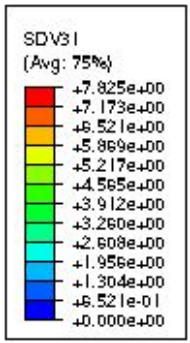

SDV 31 is accumulated
Plastic Sllip

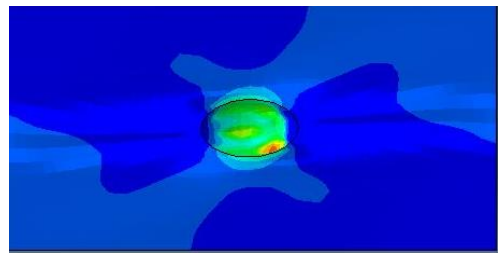

(a)  Hydrogen = 0%
Equivalent Strain = 0.43

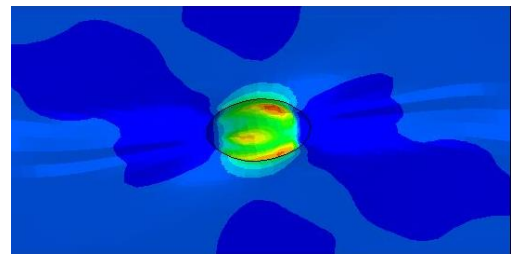

(b) Hydrogen = 3.85%
Equivalent Strain = 0.43

**Figure 9: Contour Plots showing cumulative slip activity with (a) Hydrogen Concentration = 0% and (b) Hydrogen Concentration = 3.85% Equivalent Strain = 0.3, Triaxial State = 1 and initial void fraction = 0.001**

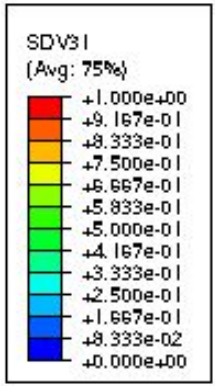 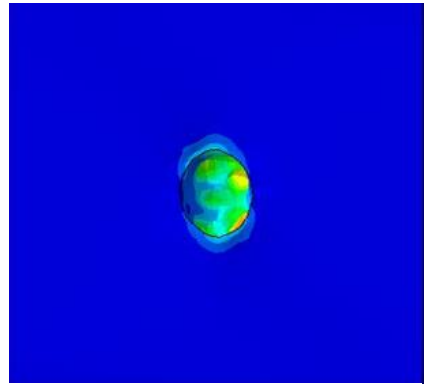 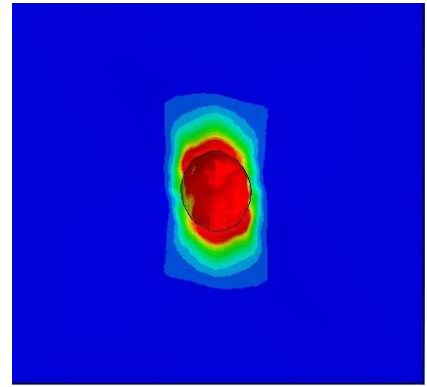

SDV 31 is accumulated
Plastic Sllip

(a)  Hydrogen = 0%
Equivalent Strain = 0.048

(b) Hydrogen = 3.85%
Equivalent Strain = 0.048

**Figure 10: Contour Plots showing cumulative slip activity with (a) Hydrogen Concentration = 0% and (b) Hydrogen Concentration = 3.85% Equivalent Strain = 0.3, Triaxial State = 3 and initial void fraction = 0.001**

Figure 11 and Figure 12 show that an increased concentration of hydrogen increases the equivalent stresses for triaxialities 1/3 and 3. Higher stresses were required for the onset of plastic deformation explained by the effect of hydrogen in locking deformation sources as previously discussed [32].

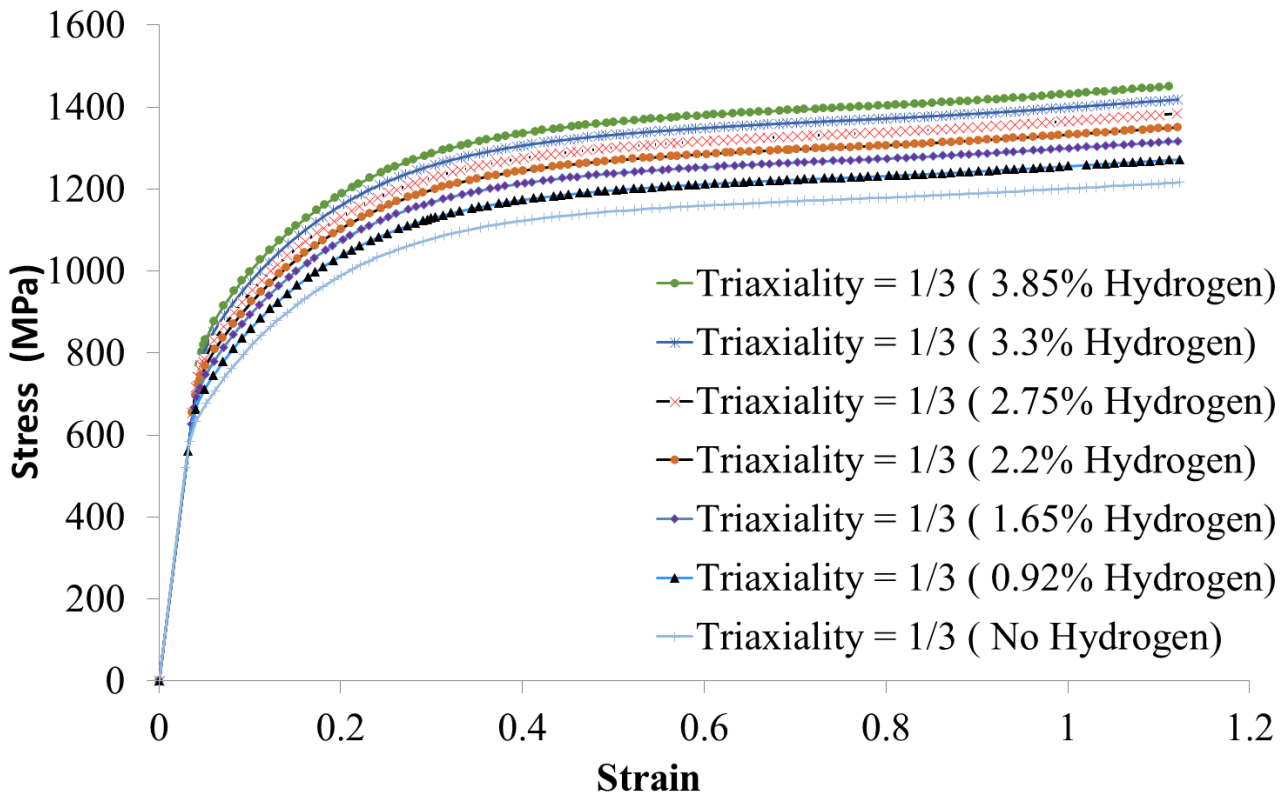

**Figure 11: Stress – Strain response at different Hydrogen Concentrations for triaxiality = 1/3 (initial void fraction = 0.001)**

Figure 13, Figure 14 and Figure 15 show that void sizes increase slowly initially and then exponentially for all triaxialities and hydrogen concentrations. The initial "slow" increase in nominal void size represents

void growth and the exponential increase represents void coalescence. For triaxiality =1, the exponential increase in void size was observed that at higher strain values when compared with samples with triaxiality =3. For samples with triaxiality = 1, it was also observed that as the concentration of hydrogen increased, the rate of increase in void fraction became higher indicating that hydrogen enhanced void growth (see Figure 13). For various hydrogen concentrations, there was also a reverse effect of hydrogen at lower strain values (see Figure 14). Hydrogen initially inhibits void growth, then beyond a strain value of approximately 0.05, hydrogen is found to promote void growth. For triaxiality =3, the void growth reduced with increasing hydrogen concentrations irrespective of strain values (see Figure 15), i.e. an increase in hydrogen concentration resulted in a reduction in the void growth.

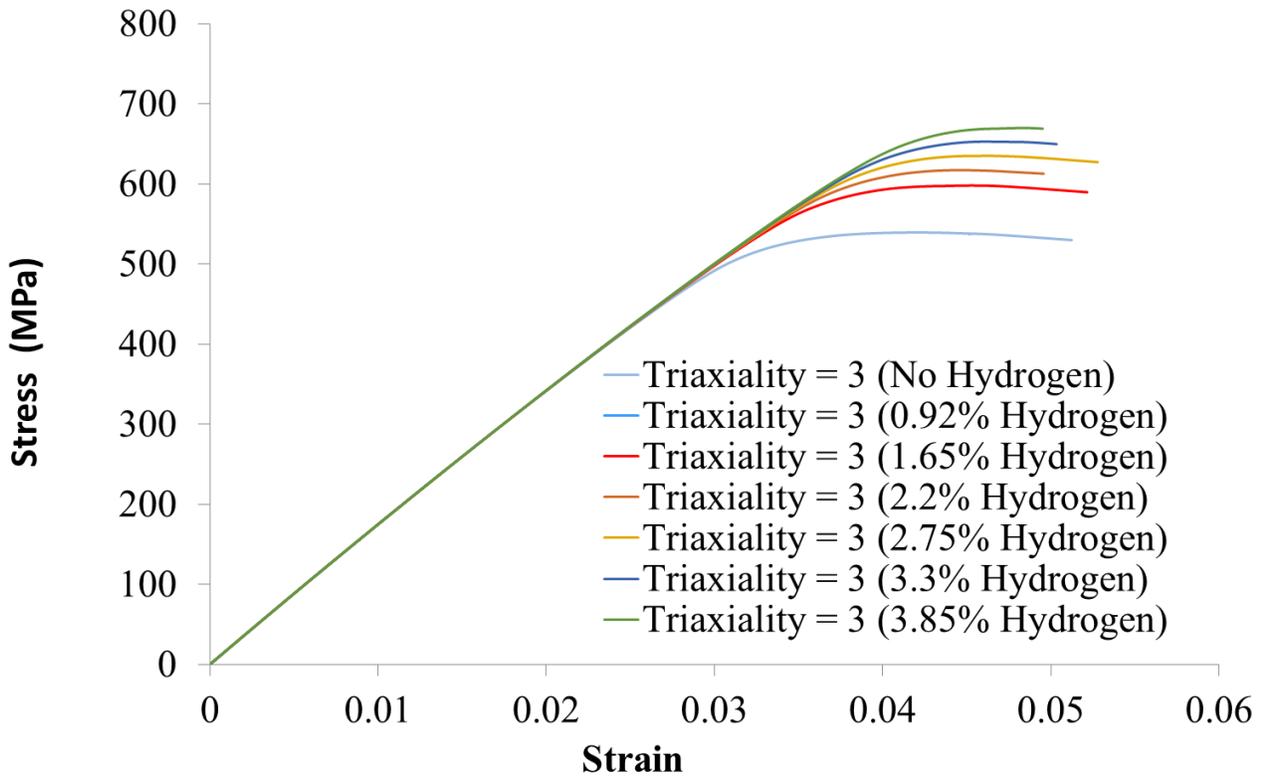

**Figure 12: Stress – Strain response at different Hydrogen Concentrations for Triaxial State = 3 (initial void fraction = 0.001)**

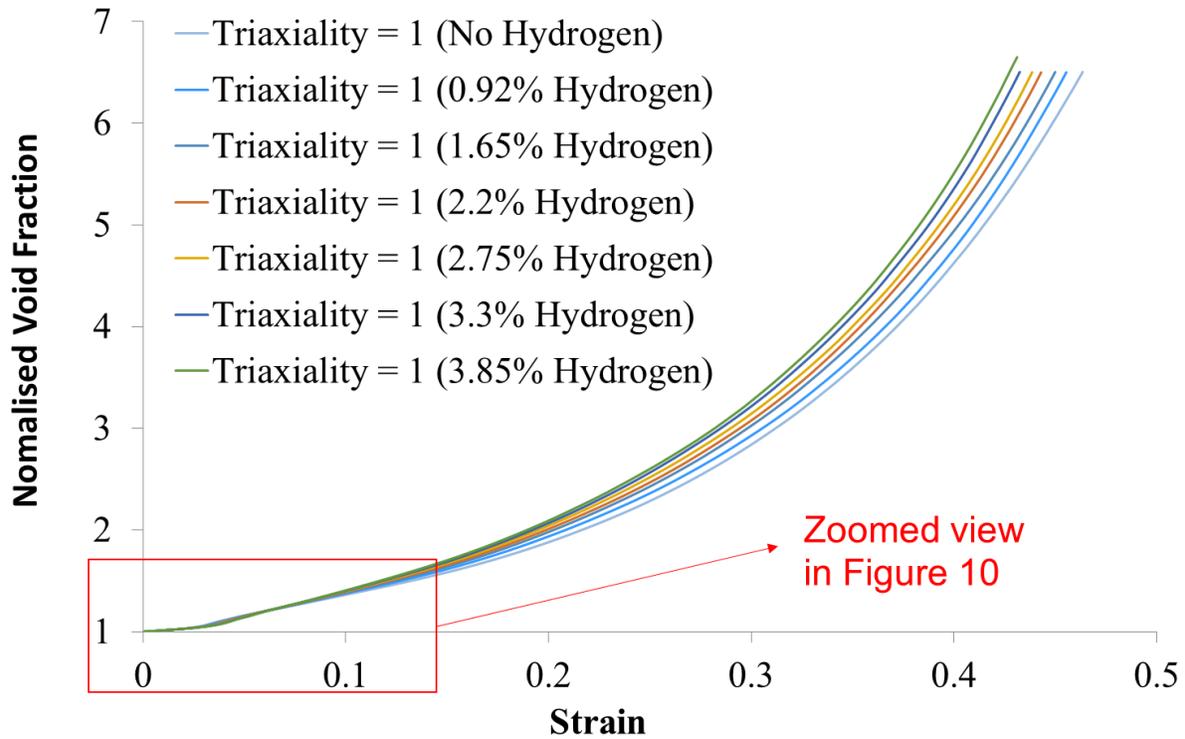

**Figure 13: Nomalised Void Fraction – Strain response at different Hydrogen Concentrations for Triaxial State = 1 (initial void fraction = 0.001)**

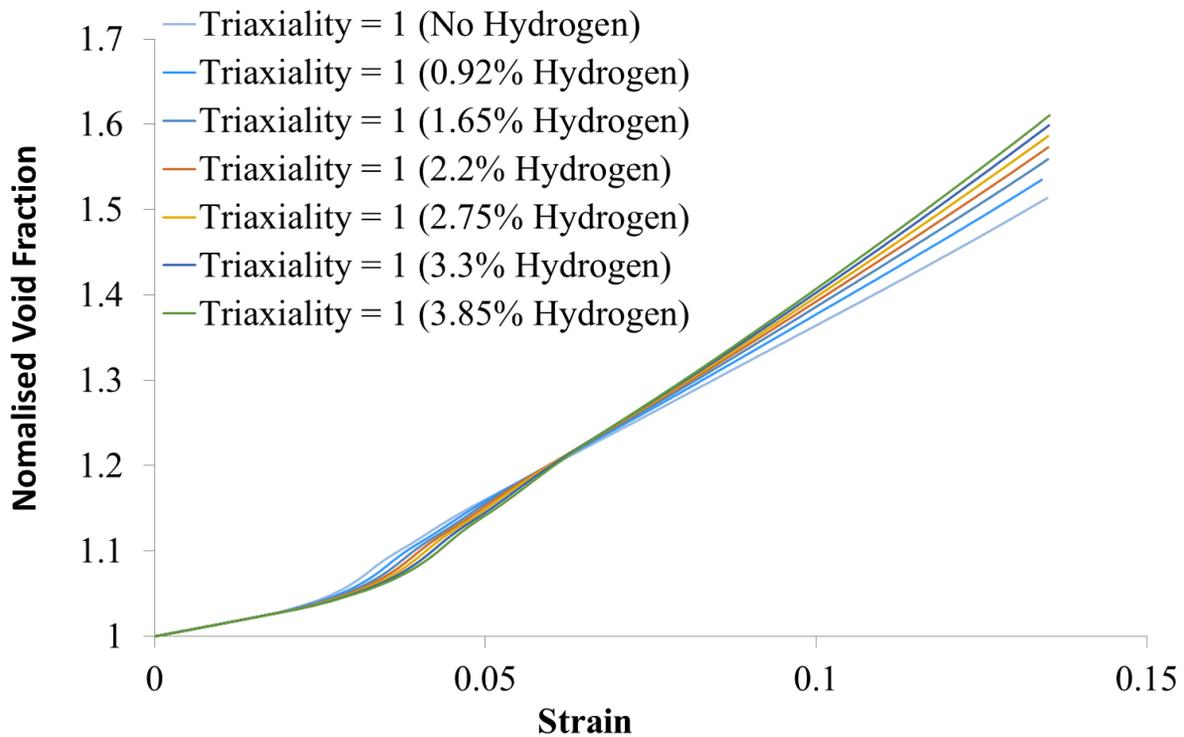

**Figure 14: Nomalised Void Fraction – Strain response at different Hydrogen Concentrations for Triaxial State = 1 (initial void fraction = 0.001) – zoomed view**

The trends shown in the figures above are consistent with the findings of Ahn et al [31] who noted that the effect of hydrogen in promoting void growth and coalescence becomes more pronounced at higher equivalent strain values. This is due to a large increase in number of traps and increased dislocation density around the void which soften the material bounding the void (see Figure **7** and

Figure **8**). The effect of hydrogen on void growth also decreases as triaxiality increases and was reversed at high triaxialities. At higher triaxialities, smaller strain values are experienced prior to coalescence so there are less microstructural defects in the vicinity of the void. As discussed in section 3, the effect of hydrogen atoms "pinning" dislocations is also a valid HELP phenomenon which has been observed experimentally so it is inferred to be more prominent under high triaxiality conditions. At triaxiality =3, the equivalent strains are lower, so hydrogen atoms hinder the motion of dislocations at the region surrounding the void and this is used to explain the observed reduction in void growth.

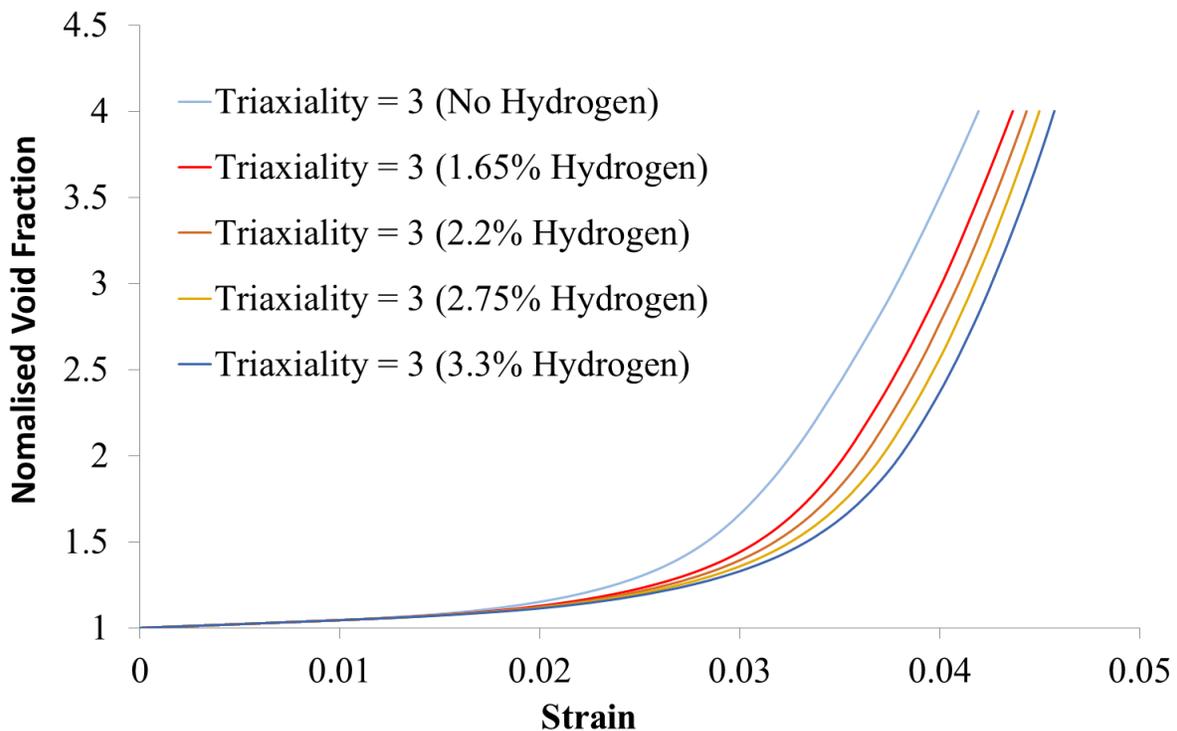

**Figure 15: Nomalised Void Fraction – Strain response at different Hydrogen Concentrations for Triaxial State = 3 (initial void fraction = 0.001)**

## 6. CONCLUSIONS

1. Hydrogen was observed to increase the equivalent stresses and hardening response for various triaxialities considered. Equivalent stresses were also found to be increased with increasing hydrogen concentration.
2. Higher void growth was observed for samples at higher triaxialities for both hydrogen containing and hydrogen free conditions. For all triaxialities considered, the void fraction is observed to increase exponentially, i.e. first increase slowly and then very rapidly.
3. For low triaxialities, the presence of hydrogen increased void growth and reduce the strain required for void coalescence. This is explained in terms of increase in microstructural defects introduced around the embedded voids caused by high strain values experienced prior to void coalescence.
4. For high triaxialities, the presence of hydrogen was observed to reduce void growth and increase the strain required for void coalescence. This is explained in terms of a dislocation "pinning effect" introduced by atomic hydrogen especially in the area surrounding the void.

## ACKNOWLEDGEMENTS

The authors are thankful to the University of Aberdeen and Apache North Sea for their support.